\documentclass[aps,prl,preprint,nopacs,superscriptaddress]{revtex4}

\usepackage{amsmath}
\usepackage{amssymb}
\usepackage{graphicx}
\usepackage{hyperref}
\usepackage[utf8]{inputenc}
\usepackage{bbm} % \mathbbmss % https://tex.stackexchange.com/questions/3492/how-do-i-make-a-blackboard-1
\usepackage{etoolbox}
\usepackage{array}

\newcommand{\beq}{\begin{equation}}
\newcommand{\eeq}{\end{equation}}
\newcommand{\beqs}{\begin{equation*}}
\newcommand{\eeqs}{\end{equation*}}

\newcommand{\ab}{{\it ab initio}}
\newcommand{\Ab}{{\it Ab initio}}
\newcommand{\exda}{Extended Data Fig. }

\newcommand{\exta}{Extended Data Table }

\newcommand{\cpana}{{\bf a}}
\newcommand{\cpanb}{{\bf b}}
\newcommand{\cpanc}{{\bf c}}
\newcommand{\cpand}{{\bf d}}
\newcommand{\cpane}{{\bf e}}

\begin{document}

% 20220116 Kludge to get left-justified section headings
\makeatletter
\def\section{%
  \@startsection
    {section}%
    {1}%
    {\z@}%
    {0.8cm \@plus1ex \@minus .2ex}%
    {0.5cm}%
    {%
      \normalfont\raggedright\small\bfseries
      %\normalfont\small\bfseries
    }%
}%
\makeatother

\title{Synthesis of a semimetallic Weyl ferromagnet\\ with point Fermi surface}

\author{Ilya Belopolski\footnote{These authors contributed equally to this work.}} \email{ilya.belopolski@riken.jp}
\affiliation{RIKEN Center for Emergent Matter Science (CEMS), Wako, Saitama 351-0198, Japan}

\author{Ryota Watanabe$^*$}
\affiliation{Department of Applied Physics and Quantum Phase Electronics Center (QPEC), University of Tokyo, Tokyo 113-8656, Japan}

\author{Yuki Sato}
\affiliation{RIKEN Center for Emergent Matter Science (CEMS), Wako, Saitama 351-0198, Japan}

\author{Ryutaro Yoshimi}
\affiliation{RIKEN Center for Emergent Matter Science (CEMS), Wako, Saitama 351-0198, Japan}

\author{Minoru Kawamura}
\affiliation{RIKEN Center for Emergent Matter Science (CEMS), Wako, Saitama 351-0198, Japan}

\author{Soma Nagahama}
\affiliation{Department of Applied Physics and Quantum Phase Electronics Center (QPEC), University of Tokyo, Tokyo 113-8656, Japan}

\author{Yilin Zhao}
\affiliation{Division of Physics and Applied Physics, School of Physical and Mathematical Sciences, Nanyang Technological University, 21 Nanyang Link, 637371, Singapore}

\author{Sen Shao}
\affiliation{Division of Physics and Applied Physics, School of Physical and Mathematical Sciences, Nanyang Technological University, 21 Nanyang Link, 637371, Singapore}

\author{Yuanjun Jin}
\affiliation{Division of Physics and Applied Physics, School of Physical and Mathematical Sciences, Nanyang Technological University, 21 Nanyang Link, 637371, Singapore}

\author{Yoshihiro Kato}
\affiliation{Department of Applied Physics and Quantum Phase Electronics Center (QPEC), University of Tokyo, Tokyo 113-8656, Japan}

\author{Yoshihiro Okamura}
\affiliation{Department of Applied Physics and Quantum Phase Electronics Center (QPEC), University of Tokyo, Tokyo 113-8656, Japan}

\author{Xiao-Xiao Zhang}
\affiliation{Wuhan National High Magnetic Field Center and School of Physics, Huazhong University of Science and Technology, Wuhan 430074, China}
\affiliation{RIKEN Center for Emergent Matter Science (CEMS), Wako, Saitama 351-0198, Japan}

\author{Yukako Fujishiro}
\affiliation{RIKEN Center for Emergent Matter Science (CEMS), Wako, Saitama 351-0198, Japan}
\affiliation{RIKEN Cluster for Pioneering Research (CPR), Wako, Saitama 351-0198, Japan}

\author{Youtarou Takahashi}
\affiliation{Department of Applied Physics and Quantum Phase Electronics Center (QPEC), University of Tokyo, Tokyo 113-8656, Japan}
\affiliation{RIKEN Center for Emergent Matter Science (CEMS), Wako, Saitama 351-0198, Japan}

\author{Max Hirschberger}
\affiliation{RIKEN Center for Emergent Matter Science (CEMS), Wako, Saitama 351-0198, Japan}
\affiliation{Department of Applied Physics and Quantum Phase Electronics Center (QPEC), University of Tokyo, Tokyo 113-8656, Japan}

\author{Atsushi Tsukazaki}
\affiliation{Institute for Materials Research, Tohoku University, Sendai, Miyagi 980-8577, Japan}

\author{Kei S. Takahashi}
\affiliation{RIKEN Center for Emergent Matter Science (CEMS), Wako, Saitama 351-0198, Japan}

\author{Ching-Kai Chiu}
\affiliation{RIKEN Interdisciplinary Theoretical and Mathematical Sciences (iTHEMS), Wako, Saitama 351-0198, Japan}

\author{Guoqing Chang}
\affiliation{Division of Physics and Applied Physics, School of Physical and Mathematical Sciences, Nanyang Technological University, 21 Nanyang Link, 637371, Singapore}

\author{Masashi Kawasaki}
\affiliation{RIKEN Center for Emergent Matter Science (CEMS), Wako, Saitama 351-0198, Japan}
\affiliation{Department of Applied Physics and Quantum Phase Electronics Center (QPEC), University of Tokyo, Tokyo 113-8656, Japan}

\author{Naoto Nagaosa}
\affiliation{Fundamental Quantum Science Program, TRIP Headquarters, RIKEN, Wako, Saitama 351-0198, Japan}
\affiliation{RIKEN Center for Emergent Matter Science (CEMS), Wako, Saitama 351-0198, Japan}

\author{Yoshinori Tokura} \email{tokura@riken.jp}
\affiliation{RIKEN Center for Emergent Matter Science (CEMS), Wako, Saitama 351-0198, Japan}
\affiliation{Department of Applied Physics and Quantum Phase Electronics Center (QPEC), University of Tokyo, Tokyo 113-8656, Japan}
\affiliation{Tokyo College, University of Tokyo, Tokyo 113-8656, Japan}


\begin{abstract}
Quantum materials governed by emergent topological fermions have become a cornerstone of physics. Dirac fermions in graphene form the basis for moir\'e quantum matter, and Dirac fermions in magnetic topological insulators enabled the discovery of the quantum anomalous Hall effect \cite{news_Anirban,CommentQuantum_Tokura, ReviewMagTopo_Tokura}. In contrast, there are few materials whose electromagnetic response is dominated by emergent Weyl fermions \cite{RMPWeylDirac_Armitage,HighFold_NatRevMat,ReviewMagTopo_BernevigFelser}. Nearly all known Weyl materials are overwhelmingly metallic, and are largely governed by irrelevant, conventional electrons. Here we theoretically predict and experimentally observe a semimetallic Weyl ferromagnet in van der Waals (Cr,Bi)$_2$Te$_3$. In transport, we find a record bulk anomalous Hall angle $> 0.5$ along with non-metallic conductivity, a regime sharply distinct from conventional ferromagnets. Together with symmetry analysis, our data suggest a semimetallic Fermi surface composed of two Weyl points, with a giant separation $> 75$\% of the linear dimension of the bulk Brillouin zone, and no other electronic states. Using state-of-the-art crystal synthesis techniques, we widely tune the electronic structure, allowing us to annihilate the Weyl state and visualize a unique topological phase diagram exhibiting broad Chern insulating, Weyl semimetallic and magnetic semiconducting regions. Our observation of a semimetallic Weyl ferromagnet offers an avenue toward novel correlated states and non-linear phenomena, as well as zero-magnetic-field Weyl spintronic and optical devices.
\end{abstract}

\date{\today}
\maketitle

Semiconductors and semimetals are materials where electrons exhibit behavior between that in insulators and metals. While semiconductors form the basis for modern electronics, and semimetals have been at the frontier of quantum physics for two decades, surprisingly, the interplay between these two fundamental phases of matter remains little explored \cite{news_Anirban,ReviewMagSemi_Ohno,CommentQuantum_Tokura, ReviewMagTopo_Tokura}. Here we report a magnetic Weyl semimetal achieved by chemical engineering of a semiconductor. Our measurements suggest that the quantum electronic structure of our material, bismuth telluride doped with chromium or (Cr,Bi)$_2$Te$_3$, exhibits a Fermi surface composed entirely of Weyl points. This places our material in a unique limit of both vanishing carrier concentration and vanishing energy gap. Only a few quantum materials are known in this extreme semimetallic regime \cite{EuCd2As2_Semi_Akrap}. Typical semimetals have a gapless electronic structure with a finite concentration of charge carriers. Established Weyl materials are semimetals in this weaker sense, hosting substantial irrelevant charge carriers that coexist with the Weyl fermions and obscure their unique properties \cite{RMPWeylDirac_Armitage,HighFold_NatRevMat,ReviewMagTopo_BernevigFelser}. Trivial, metallic electrons are intrinsic to the TaAs family of inversion-breaking Weyl materials; the RhSi family of topological chiral crystals; ferromagnetic Weyl kagome Co$_3$Sn$_2$S$_2$; and linked Weyl ring Co$_2$MnGa \cite{RMPWeylDirac_Armitage,HighFold_NatRevMat,ReviewMagTopo_BernevigFelser,Co2MnGa_me2}. By contrast, a material with only Weyl points at the Fermi level will exhibit electromagnetic response dominated by Weyl physics. Such a semimetallic Weyl material is of urgent interest and expected to support novel device functionality, including topological transistors \cite{HighFold_NatRevMat, Na3Bi_phasetransition_Edmonds}, giant photovoltaic effects \cite{Burch_PhotovoltaicWeyl}, Weyl spin-charge switches \cite{Mn3Sn_films_Nakatsuji}, non-linear terahertz components \cite{EfficientTerahertz_ArmitageMatsunaga,TerahertzDirac_OkaWang}, Majorana-Fermi arcs \cite{Beenakker_WeylMajorana_2017}, energy-harvesting thermoelectrics \cite{Fe3Ga_Nakatsuji, EnergyHarvesting_Nakatsuji_2019} and topological catalysts \cite{FeslerRao_WeylCatalysts_2017}. A Fermi surface of only Weyl points is further expected to fail to screen Coulomb repulsion, producing a logarithmically divergent correction to Fermi velocities \cite{RMP_CastroNeto_2012,Guinea_Geim_DiracReshaped_2011,WitczakKrempa_InteractingWeyl_2014, Bergholtz_InteractingWeyl_2018}, and emergent Lorentz invariance driven by interactions \cite{Isobe_WeylTilt_2016}.

To realize a semimetallic Weyl phase of matter, it is natural to start from a semiconductor, without irrelevant metallic bands anywhere in the Brillouin zone. Chemical substitution can then introduce ferromagnetism, producing a Zeeman splitting which can drive the formation of emergent Weyl fermions \cite{ReviewMagSemi_Ohno}. Using a semiconductor with large spin-orbit coupling (SOC) may also give wider access to the topological phase space (Fig. \ref{Fig1}{\bf a}). Such a scheme was at the core of early proposals for Weyl semimetals \cite{HgCdMnTe_Bulmash,HgCr2Se4Weyl_Xu,ARCMP_me}. Notably the Burkov-Balents multilayer constructs a Weyl semimetal from a stack of ferromagnetic topological insulator and trivial insulator layers \cite{TINI_BurkovBalents, TINI_AtomicChain_me,Gapless_Murakami}. Despite considerable interest, these proposals have not yet been experimentally realized. Bi$_2$Te$_3$ is a topological semiconductor which has already been ferromagnetically doped to produce a quantum anomalous Hall (QAH) state \cite{ReviewMagTopo_Tokura}. To produce Weyl fermions in Bi$_2$Te$_3$, it is necessary to break time-reversal symmetry by introducing a ferromagnetic exchange splitting $J$, sending the material from the topological insulating state towards a semimetallic Weyl state. It may also be beneficial to increase the mass gap $m$ by reducing the SOC. Interestingly, both can be achieved by chemically substituting Cr into pristine Bi$_2$Te$_3$ (Fig. \ref{Fig1}{\bf b,c}). 

We synthesized films of Bi$_2$Te$_3$ doped with Cr, with an additional low level of Sb co-doping to achieve charge neutrality (Composition A, Table \ref{comptable}). Note that this composition includes significantly higher Cr content than is typically used to produce a QAH state \cite{ReviewMagTopo_Tokura,QAHStandard_Okazaki_2021}. In transport we first examine the two-dimensional Hall sheet conductivity $\sigma^{2 {\rm D}}_{xy} (B)$ of such a (Cr,Bi,Sb)$_2$Te$_3$ film, of thickness $h = 100$ nm (Fig. \ref{Fig1} {\bf d}). We obtain the anomalous Hall effect (AHE) $\sigma^{2 {\rm D}}_{\rm AHE}$ at $T = 2$ K by extrapolating the high-magnetic-field $\sigma^{2 {\rm D}}_{xy} (B)$ back to zero field. We find $\sigma^{2 {\rm D}}_{\rm AHE} \sim 24\ e^2/h$, written in units of the conductance quantum $e^2/h \sim 3.9 \times 10^{-5}$ $\Omega^{-1}$. This result is somewhat unexpected because topological insulators with ferromagnetism typically exhibit a QAH generated by magnetically-gapped Dirac cone surface states. In this well-known scenario $\sigma^{2 {\rm D}}_{\rm AHE}$ can never exceed $1\ e^2/h$. Our observation of $\sigma^{2 {\rm D}}_{\rm AHE} \gg 1\ e^2/h$, suggests that the observed Hall effect is instead driven by bulk Berry curvature. Since Cr doped into Bi$_2$Te$_3$ introduces ferromagnetism and reduces SOC, it naturally drives a transition from a topological insulator to a Weyl semimetal \cite{Gapless_Murakami, TINI_BurkovBalents}. To theoretically understand this phase transition, we can consider a minimal $k \cdot p$ model for Bi$_2$Te$_3$ describing the electronic structure in the vicinity of the bulk $\Gamma$ point \cite{HaijunZhang_Bi2Te3, ChaoXingLiu_Bi2Te3}. In the presence of ferromagnetism, to linear order in $k$, the low-energy electronic structure is governed by,
\beq
h({\bf k})_{\rm (Cr,Bi)_2Te_3} = v_{\rm F}({\bf k} \cdot \sigma) \tau_x + m \tau_z + J \sigma_z %J \hat{n} \cdot \sigma
\label{hamBT}
\eeq

\noindent where ${\bf k}$ is the crystal momentum, $\sigma$ acts in spin space, $\tau$ acts in orbital space, $m$ sets the mass gap and $J$ sets the ferromagnetic exchange splitting. Suppose that the mass gap is positive far from $\Gamma$, at all other time-reversal invariant momenta. Then pristine Bi$_2$Te$_3$ has $m < 0$, $J = 0$, giving a band inversion at $\Gamma$ which produces a three-dimensional $\mathbb{Z}_2$ topological insulator. The case $m > 0$ would correspond to a trivial insulator. Ferromagnetic (Cr,Bi)$_2$Te$_3$ is a Weyl semimetal whenever the exchange splitting $J$ overwhelms the mass gap $m$, with exactly two Weyl points along the $k_z$ axis at $k_z^2 = J^2 - m^2$. Crucially, Bi$_2$Te$_3$ has no irrelevant electronic states near the Fermi level elsewhere in the bulk Brillouin zone. As a result, if the system is tuned to charge neutrality, the Fermi surface consists of only two Weyl points. Furthermore, recall that the AHE in a Weyl semimetal is proportional to the separation of the Weyl points in momentum space $\sigma_{xy}^{\textrm{3D}} = (\Delta k / 2 \pi) (e^2 / h)$ \cite{AHEWeyl_Burkov_2014}. In our case, this implies that we can extract the Weyl point separation directly from the measured Hall conductivity, when the system is tuned to charge neutrality. For `Composition A', we find a separation of $\sim 75\%$ of the bulk Brillouin zone in the $k_z$ direction (Fig. \ref{Fig1}{\bf c}, right axis), larger than any Weyl material studied to date.

To more deeply understand this large Weyl point separation, we performed \textit{ab initio} calculations of (Cr,Bi)$_2$Te$_3$. We observe two Weyl points at the Fermi level, with a large separation of $\sim 90\%$ of the Brillouin zone, broadly consistent with our experiment (Fig. \ref{Fig1}{\bf c}). Examining the measured longitudinal conductivity, we find a relatively small value, $\sigma^{2 {\rm D}}_{xx} \sim 47$ $e^2/h$, giving a Hall angle, $\sigma_{\rm AHE} / \sigma_{xx} \sim 0.51$, again larger than any Weyl material studied to date. We can naturally understand the giant bulk Hall angle given the simplicity of our system, which exhibits the minimum number of Weyl points, with no irrelevant electronic states at the Fermi level and chemical potential tuned to charge neutrality. This transport data, theoretical analysis and \textit{ab initio} calculation together suggest the observation of a semimetallic Weyl ferromagnet in (Cr,Sb,Bi)$_2$Te$_3$. Despite numerous theoretical proposals and experimental attempts over the course of more than a decade, a magnetic Weyl semimetal has never before been produced from a semiconductor \cite{TINI_BurkovBalents,TINI_AtomicChain_me,HgCdMnTe_Bulmash,HgCr2Se4Weyl_Xu}. We can simplify the film composition by using Cr alone to introduce ferromagnetism, suppress SOC and bring the material toward charge neutrality. In (Cr,Bi)$_2$Te$_3$ `Composition B', we again observe $\sigma_{xy} \gg 1\ e^2/h$, with Weyl point separation $\sim 77\%$ (Fig. \ref{Fig1}{\bf e}). The large Cr level in our semimetallic Weyl films further drives a large Curie temperature $T_{\rm C} \sim 150$ K (Fig. \ref{Fig1}{\bf e}, inset) and enhanced coercive field, indicating a robust ferromagnetic Weyl phase.

%\begin{spacing}{2}
\begin{table}
\begin{center}
\begin{tabular}{>{\centering\arraybackslash}p{2.7cm} >{\centering\arraybackslash}p{2cm} >{\centering\arraybackslash}p{2cm} >{\centering\arraybackslash}p{2cm} >{\centering\arraybackslash}p{2cm} >{\centering\arraybackslash}p{2cm}}
Composition & Cr & In & Sb & Bi & $\Delta k_{\textrm{Weyl}}$ \\
\hline	
A & 0.24 & 0 & 0.20 & 0.56 & 75\% \\
B & 0.30 & 0 & 0 & 0.70 & 77\% \\ %Sample 2268
C & 0.13 & 0.02 & 0.57 & 0.28 & 15\% \\
D & 0.05 & 0 & 0.64 & 0.31 & Chern \\
E & 0.27 & 0.06 & 0.45 & 0.22 & trivial \\
calc. & 0.33 & 0 & 0 & 0.67 & 90\% \\
calc. & 0.33 & 0 & 0.67 & 0 & 11\% \\
\end{tabular}
\caption{{\bf Weyl compositions.} Representative samples examined in the present work, all based on (Cr,Bi)$_2$Te$_3$. The values are written as (Cr$_x$In$_y$Bi$_z$Sb$_{1-x-y-z}$)$_2$Te$_3$. An additional 34 measured devices are marked in Fig. \ref{Fig3}b.}
\label{comptable}
\end{center}
\end{table}
%\end{spacing}

To more carefully demonstrate a semimetallic Weyl ferromagnet, we examine the transport properties of our films as a function of thickness. Here we additionally co-dope with a low level of indium, In, which predominantly serves to reduce the SOC. This allows us to use a lower Cr level and consider two compositions which are chemically similar, but with (In,Cr,Bi,Sb)$_2$Te$_3$ `Composition C' targeting a semimetallic Weyl state, and a nearby (Cr,Bi,Sb)$_2$Te$_3$ `Composition D' targeting a Chern insulator (Table \ref{comptable}). For both compositions, we maintain approximate charge neutrality at a ratio Bi : Sb $\sim$ 1 : 2. We observe that the series of films in `Composition C' exhibit a Hall sheet conductance which scales linearly with the film thickness, indicating an AHE driven by bulk Berry curvature (Fig. \ref{Fig2}{\bf a},{\bf c}). By contrast, the series `Composition D' exhibits an AHE quantized to $\pm 1\ e^2 / h$ regardless of the film thickness, indicating that the Hall response is driven by the top and bottom surfaces of the film (Fig. \ref{Fig2}{\bf b}). We can again extract the Weyl point separation for our (In,Cr,Bi,Sb)$_2$Te$_3$ Weyl films, where we observe a modest $\Delta k_{\rm Weyl} \sim 15\%$ of the Brillouin zone, consistent with the lower Cr level and consequently weaker ferromagnetism. As an additional check, we examine a more heavily In-doped `Composition E', which shows no appreciable AHE, suggesting the absence of Berry curvature and indicating a topologically-trivial magnetic semiconductor (\exda \ref{trivial}). We also perform \ab\ calculations for Cr-substituted Sb$_2$Te$_3$, which more closely captures the high Sb level of `Composition C'. We obtain $\Delta k_{\rm Weyl} \sim 11\%$ (Fig. \ref{Fig1}{\bf d},{\bf e}), roughly in agreement with transport. We further explore the Faraday rotation of `Composition C' in THz spectroscopy, which is directly sensitive to the bulk topological nature of the system (\exda \ref{THz}). We observe an optical response consistent with DC transport, providing independent experimental evidence for a semimetallic Weyl ferromagnet.

We can more deeply explore the Weyl state via the chiral anomaly, which gives rise to a negative longitudinal magnetoresistance under parallel electric and magnetic fields \cite{ReviewChiralAnomaly_Ong,ARCMP_me}. We measure the longitudinal resistivity under in-plane magnetic field $B$, as a function of the angle $\theta$ between the applied current $I$ and $B$, for films with varying Cr and In levels (Fig. \ref{Fig3}{\bf a}). We find that the Chern insulator `Composition D' exhibits a maximum in the resistivity under $I \  || \  B$, which is the typical magnetoresistance anisotropy observed in conventional magnets, as well as topological insulators \cite{CrBiSbTe_AMR_Samarth}. We observe a similar behavior in `Composition E', targeting a topologically-trivial magnetic semiconducting state. By contrast, the Weyl `Composition C' exhibits a suppression of the resistivity under parallel $I$ and $B$, suggesting a contribution from the chiral anomaly. To further explore the chiral anomaly, we leverage (In,Cr) doping to precisely vary the ferromagnetism and SOC throughout the topological phase diagram, across a library of 37 films (including A, C, D and E; Fig. \ref{Fig3}{\bf b}). At low (In,Cr), corresponding to a Chern insulator, and at high (In,Cr), corresponding to a trivial magnetic semiconductor, we observe a conventional, positive magnetoresistance anisotropy (color map in Fig. \ref{Fig3}{\bf b} obtained from $\Delta \rho_{xx}(\theta = 0^{\circ})$ measured at each black dot). At intermediate Cr levels we consistently observe a negative magnetoresistance anisotropy, suggesting a chiral anomaly driven by the Weyl phase.

To gain further insight into this region of the composition space, we consider the conceptual phase diagram proposed by Murakami \cite{Gapless_Murakami}. This phase diagram maps out the electronic state as a function of symmetry breaking and the mass gap. For the case of time-reversal symmetry breaking, we can draw a magnetic Murakami diagram plotted as a function of ferromagnetic exchange splitting and the mass gap (Fig. \ref{Fig1}{\bf a}, Fig. \ref{Fig3}{\bf c}). This diagram includes Weyl semimetal phases separating Chern and trivial insulator phases. To relate our color map of the magnetoresistance anisotropy to the Murakami diagram, again recall that Cr doping both introduces ferromagnetism and reduces SOC, while In doping only reduces SOC. We can therefore convert the (In,Cr) axes to (magnetism, SOC) axes via a linear coordinate transformation, in a way which associates the positive and negative magnetoresistance regions with the Murakami phase regions. Under a reasonable coordinate transformation, we find that the negative magnetoresistance region coincides well with the Weyl phase, while the positive regions coincide well with the Chern insulator and magnetic semiconductor regions. These regions are also consistent with the bulk scaling of the AHE observed for Weyl `Composition C', surface scaling of the QAH in `Composition D' and absence of AHE in semiconductor `Composition E' (Fig. \ref{Fig2}{\bf c}). Our systematic measurements indicate the observation of a highly-tunable chiral anomaly associated with the Weyl state, and suggest that we can access large Weyl semimetallic, Chern insulator and magnetic semiconductor regions of the Murakami phase space. In this way, we visualize the foundational Murakami phase diagram for the first time.

To more deeply understand (Cr,Bi)$_2$Te$_3$ we consider the well-known Burkov-Balents multilayer, which offers the simplest theoretical recipe for a magnetic Weyl semimetal \cite{TINI_BurkovBalents}. This model consists of layers of a topological insulator similar to Bi$_2$Te$_3$, layers of a trivial insulator and ferromagnetism. The multilayer hosts a stack of coupled two-dimensional Dirac cones, which hybridize with one another to produce an emergent three-dimensional electronic structure hosting a single pair of Weyl points (\exda \ref{BBcartoon}). The Hamiltonian is,
\beq
h({\bf k})_{\rm BB} = v_{\rm F} (-\sigma_y k_x + \sigma_x k_y) \tau_z + J \sigma_z + (t - u) \tau_x + k_z (ta + ub)\tau_y
\label{hamBB}
\eeq

\noindent Here the first term captures the two-dimensional Dirac cone living at each interface within the stack, $t$ represents the hopping amplitude across the topological insulator layer, $u$ represents the hopping amplitude across the trivial insulator layer, $J$ sets the exchange splitting, $a$ is the thickness of the topological layer and $b$ is the thickness of the trivial layer. Despite considerable interest, it has been challenging to realize the Burkov-Balents proposal directly \cite{TINI_AtomicChain_me,HighChern_CuiZuChang,MnBi2Te4_multi_MacDonald}. At the same time, it is intriguing to note that (Cr,Bi)$_2$Te$_3$ is also based on Bi$_2$Te$_3$ and similarly exhibits a single pair of Weyl points. This suggests a close relationship to the Burkov-Balents model. Indeed, bulk Bi$_2$Te$_3$ itself can be viewed as a stack of coupled two-dimensional Dirac cones with $t < u$. In fact, $h({\bf k})_{\rm (Cr,Bi)_2Te_3}$ (Eq. \ref{hamBT}) and $h({\bf k})_{\rm BB}$ (Eq. \ref{hamBB}) are equivalent up to a unitary transformation, and describe the same physics (see Methods). Using this equivalence, we see that the fine multilayer structure is unnecessary and homogeneous doping in (Cr,Bi)$_2$Te$_3$ achieves the same minimal Weyl electronic structure as the Burkov-Balents model.

A decade of exploration of Weyl physics in quantum matter has led to the discovery of numerous materials with unusual properties \cite{RMPWeylDirac_Armitage,HighFold_NatRevMat,ReviewMagTopo_BernevigFelser,ARCMP_me}. However, far from being semimetallic, nearly all of these materials are dominated by conventional metallic response (Fig. \ref{Fig4}{\bf a}). Our systematic transport measurements and chemical composition dependence, combined with \textit{ab initio} calculations and analytical theory, show that (Cr,Bi)$_2$Te$_3$ offers a semimetallic Weyl quantum material. In sharp contrast to existing Weyl materials, this platform is based on a semiconductor and exhibits a Weyl point Fermi surface. Examining $\rho_{xx} (T)$, we directly observe an insulating behavior above the Curie temperature, $T > T_{\textrm{C}}$, suggestive of a semiconductor, followed by a decrease in the resistivity at $T < T_{\textrm{C}}$ as the ferromagnetism drives a Weyl state (Fig. \ref{Fig4}{\bf a}). We further observe a giant Hall angle $\tan \theta_{\rm H} = \sigma_{\textrm{AHE}}/\sigma_{xx} = 0.51$ at our optimal composition, more than double existing Weyl materials, again suggesting a semimetallic Fermi surface (Fig. \ref{Fig4}{\bf b}). We also find an exceptionally large separation of Weyl points, suggesting that (Cr,Bi)$_2$Te$_3$ offers a robust and highly-tunable semimetallic Weyl platform (Fig. \ref{Fig4}{\bf c}). Lastly, we consider (Cr,Bi)$_2$Te$_3$ in the context of the unified scaling relations governing $\sigma_{\rm AHE}$ and $\sigma_{xx}$ \cite{Onoda_TransportHall_2008}. Materials typically fall in one of three regimes: a high-conductivity skew scattering regime; a low-conductivity localized hopping regime; and an intermediate regime governed by Berry curvature, with $\sigma_{xx} \sim 10^4$ to $10^6$ $\Omega^{-1} {\rm cm}^{-1}$ (Fig. \ref{Fig4}{\bf d}). Although $\sigma_{\rm AHE}$ in (Cr,Bi)$_2$Te$_3$ is produced by Berry curvature, the low $\sigma_{xx}$ drives the system outside the conventional regime of the intrinsic anomalous Hall effect, with $\sigma_{xx} \sim 10^2$ $\Omega^{-1} {\rm cm}^{-1}$. This unique behavior is again a direct consequence of electromagnetic response dominated by Weyl fermions, rather than conventional accumulation of Berry curvature in the electronic structure. The synthesis of a semimetallic Weyl quantum material, with a Fermi surface consisting only of Weyl points, offers unprecedented opportunities to incorporate emergent Weyl fermions into device architectures. Such a semimetallic system also offers opportunities to investigate breakdown of the Fermi liquid picture and renormalization of Fermi velocities \cite{RMP_CastroNeto_2012,Guinea_Geim_DiracReshaped_2011,WitczakKrempa_InteractingWeyl_2014, Bergholtz_InteractingWeyl_2018}, arising from the interplay of emergent Weyl fermions and electron-electron interactions.

%\bibliography{/Users/i99999/Documents/master_bib.bib}{}
%\bibliographystyle{/Users/i99999/Documents/Science.bst}

\section*{\uppercase{Methods}}

{\it Equivalence to the Burkov-Balents theory:} We show that the low-energy effective theory describing (Cr,Bi)$_2$Te$_3$ is equivalent to the well-known model for an ideal ferromagnetic Weyl semimetal in a topological insulator multilayer (\exda \ref{BBcartoon}) \cite{TINI_BurkovBalents},
\beq
H_{\rm BB} = \sum_{i, j} \bigg[ v_{\rm F} (-\sigma_y k_x + \sigma_x k_y) \tau_z \delta_{ij} + J \sigma_z \delta_{ij} + t \tau_x \delta_{ij} - u \tau_+ \delta_{i,j+1} - u \tau_- \delta_{i,j-1} \bigg] c^{\dag}_i c_j
\eeq
Here the first term describes the topological insulator Dirac cone, the second term describes the ferromagnetic exchange splitting and the remaining terms describe the amplitudes $t$ and $u$ for the Dirac cones to hop across the topological and trivial layers. There are two bands arising from the Dirac cone and two Dirac cones per unit cell, giving a four-band model. For convenience, we define $t$ and $u$ with opposite signs. Summation over $k_x$, $k_y$ and the lattice basis is implied. In momentum space, we have,
\beq
H_{\rm BB} = \sum_{k} c^{\dag}_k h({\bf k}) c_k, \hspace{0.7cm} h({\bf k}) = v_{\rm F} (-\sigma_y k_x + \sigma_x k_y) \tau_z + J \sigma_z + (t e^{i k_z a}  + u e^{- i k_z b}) \tau_+ + h.c.
\eeq
This is equivalent to Eq. (9) in Ref. \cite{TINI_BurkovBalents}. Expanding around the $\Gamma$ point, $k_z = 0$, we obtain $h({\bf k})_{\rm BB}$ (Eq. \ref{hamBB}). Crucially, $h({\bf k})_{\rm (Cr,Bi)_2Te_3}$ and $h({\bf k})_{\rm BB}$ can be related by the unitary transformation $U = {\rm diag}(1, -i, 1, i)$,
\beq
h({\bf k})_{\rm BB} = U^{\dag} h({\bf k})_{\rm (Cr,Bi)_2Te_3} U
\eeq
Under this transformation, $m \tau_z \leftrightarrow (t-u) \tau_x$ and $v_{\rm F} k_z \sigma_z \tau_x \leftrightarrow k_z (ta + ub) \tau_y$. Intuitively, if the topological insulator layers are very thin, the system should essentially become a ferromagnetic trivial insulator. In this regime, the hopping across the topological layer dominates, $t \gg u$, so $t - u > 0$ and we see that the mass gap is trivial. By contrast, if the trivial insulator layers are very thin, the system is essentially a ferromagnetic topological insulator. This corresponds to $u \gg t$, so $t - u < 0$ and the mass gap is topological. We have shown that the two models are equivalent and homogeneously-doped (Cr,Bi)$_2$Te$_3$ can be viewed as realizing the Burkov-Balents proposal.\\

{\it Sample synthesis:} (Cr,Bi)$_2$Te$_3$, (Cr,Bi,Sb)$_2$Te$_3$ and (Cr,In,Bi,Sb)$_2$Te$_3$ films were synthesized by molecular beam epitaxy (MBE) in a system equipped with standard Knudsen cells and maintained at a base pressure of $10^{-7}$ Pa. Films were deposited on epi-ready semi-insulating InP(111)A substrates which were chemically etched by 10\% H$_2$SO$_4$, as well as BaF$_2$(111). Substrates were maintained during film synthesis at a growth temperature of $200^{\circ}$C. The nominal composition of films was determined by beam equivalent pressures of Cr, In, Bi, Sb and Te fluxes, calibrated prior to film growth. The beam pressure of Te was set at a ratio of 40:1 relative to the cation elements, to suppress Te vacancies. The typical growth rate was $1$ nm per $4$ min, as determined by low-angle Laue fringes in X-ray diffraction. To prevent deterioration in atmosphere, films were capped by 3 nm of AlO$_x$ using atomic layer deposition (ALD) at room temperature immediately after being removed from the MBE system. The actual Cr level was later calibrated via inductively coupled plasma mass spectroscopy (ICP-MS) and found to be higher than the nominal level fixed by beam equivalent pressure $x_{\rm ICP-MS} \sim 5.7 \times x_{\rm nominal}$ (\exda \ref{ICP}). Hall bar devices of typical size 100 $\mu$m $\times$ 100 $\mu$m were fabricated using standard UV photolithography with Ar ion milling. Electrodes were deposited by electron beam evaporation of Ti (5 nm) followed by Au (45 nm).\\

{\it Transport measurements:} The electrical resistance was measured using a Quantum Design Physical Property Measurement System (PPMS) at temperatures 2 K to 300 K under magnetic fields up to 9 T and an excitation current of 10 $\mu$A. For Compositions A and B, a bare film was measured, with Hall bars defined by gold wires fixed by Ag epoxy at distances of several millimeters. All other compositions were measured using lithographically patterned Hall bars with $W$ and $L$ of typical size 100 $\mu$m. The resistivity was extracted by scaling the resistance by the sample dimensions, $\rho^{2{\rm D}}_{xx} = R_{xx} W / L$, $\rho_{yx} = R_{yx}$ and $\sigma^{2{\rm D}}$ was obtained by inverting the $\rho^{2{\rm D}}$ tensor. See \exda \ref{A_rho} for the resistivity measurements used to plot Fig. \ref{Fig1}{\bf d,e}. The three-dimensional bulk resistivity and conductivity were calculated using the film thickness $h$, $\rho = \rho^{2{\rm D}} h$ and $\sigma = \sigma^{2{\rm D}}/h$. The Weyl point separation was extracted from the three-dimensional conductivity as described in the main text, which gives $\Delta k_{\rm Weyl}$ in units of ${\rm \AA}^{-1}$. The size the bulk Brillouin zone in $k_z$ is $2\pi/c$, where $c$ is the lattice constant of (Cr,Bi)$_2$Te$_3$. The Weyl point separation was then obtained as a percent of $2\pi/c$, so that 100\% corresponds to the Weyl points meeting up at the boundary of the bulk Brillouin zone.\\

{\it Density functional theory:} Calculations were performed within the density functional theory \cite{HohenbergKohn_1964} framework with the projector augmented wave basis, using the Vienna \ab\ simulation package \cite{DFT2}. The Perdew-Burke-Ernzerhof-type generalized gradient approximation \cite{DFT4} was used to describe the exchange-correlation energy. A plane wave cutoff of 300 eV was used, and a $12 \times 12 \times 6$ $k$-mesh was chosen to sample the bulk Brillouin zone. The total energies were converged to $10^{-6}$ eV. Spin-orbit coupling was taken into account self-consistently to treat relativistic effects. See \exta \ref{Ext_Struct} for crystal structures.\\

\makeatletter
\apptocmd{\thebibliography}{\global\c@NAT@ctr 40\relax}{}{}
\makeatother

\section*{\uppercase{Acknowledgements}}

I.B. acknowledges discussions on transport measurements with Max T. Birch and discussions regarding BaF$_2$ substrates with Masao Nakamura. I.B. and R.Y. acknowledge thin film synthesis by Taiga Ueda. The authors acknowledge Kaustuv Manna for transport data on Co$_2$MnGa and Chenglong Zhang for transport data on TaAs. The authors acknowledge Yaojia Wang, Shuo-Ying Yang and Mazhar N. Ali for anomalous Hall scaling data on reference materials. I.B. acknowledges discussions on Weyl physics with Lixuan Tai, Masataka Mogi, Hiroki Isobe, Tyler A. Cochran, Daniel S. Sanchez, Qiong Ma, Su-Yang Xu, Joe Checkelsky and Kenji Yasuda. The authors are grateful to all members of CEMS for discussions. This work was supported by the Japan Society for the Promotion of Science (JSPS), KAKENHI Grant No. 23H05431 (Y. Tokura), 24K17020 (Y.S.), 24H01607 (M.H.), 22H04958 (M. Kawasaki), 24H00197 (N.N.), 24H02231 (N.N.) and 23H01861 (M. Kawamura); and by the Japan Science and Technology Agency (JST) FOREST JPMJFR2238 (M.H.). N.N. was also supported by the RIKEN TRIP Initiative. C.-K.C. was supported by JST Presto Grant No. JPMJPR2357 and partially supported by JST as part of Adopting Sustainable Partnerships for Innovative Research Ecosystem (ASPIRE) Grant No. JPMJAP2318. This work was further supported by the RIKEN TRIP Initiative (Many-Body Electron Systems). Work at Nanyang Technological University was supported by the National Research Foundation, Singapore, under its Fellowship Award (NRF-NRFF13-2021-0010), the Agency for Science, Technology and Research (A*STAR) under its Manufacturing, Trade and Connectivity (MTC) Individual Research Grant (IRG) (Grant No. M23M6c0100), the Singapore Ministry of Education (MOE) Academic Research Fund Tier 3 Grant (MOE-MOET32023-0003), Singapore Ministry of Education (MOE) AcRF Tier 2 Grant (MOE-T2EP50222-0014) and the Nanyang Assistant Professorship Grant (NTU-SUG).

\section*{\uppercase{Author contributions}}

Y. Tokura and N.N. supervised the project. I.B. conceived the research. R.W. and I.B. synthesized thin films and fabricated devices with help from Y.S. and S.N. and under the guidance of R.Y., M. Kawamura, A.T., K.S.T., M. Kawasaki and Y. Tokura. Y.Z., S.S. and Y.J. performed first-principles calculations under the supervision of G.C. Y.K., Y.O., I.B. and X.X.Z. acquired and analyzed THz spectra under the guidance of Y. Takahashi, N.N. and Y. Tokura. R.W. and I.B. performed transport measurements with help from Y.S., Y.F. and M.H. C.K.C. performed the numerical calculations with I.B. and Y.S. I.B. wrote the theory under the guidance of C.K.C., N.N. and Y. Tokura. I.B. wrote the manuscript with contributions from all authors.

% \section*{\uppercase{Data availability}}

\section*{\uppercase{Competing interests}}

The authors declare no competing interests.

% \section*{\uppercase{Additional information}}

\clearpage
\begin{figure}[h]
\centering
\includegraphics[width=12cm,trim={0in 0in 0in 0in},clip]{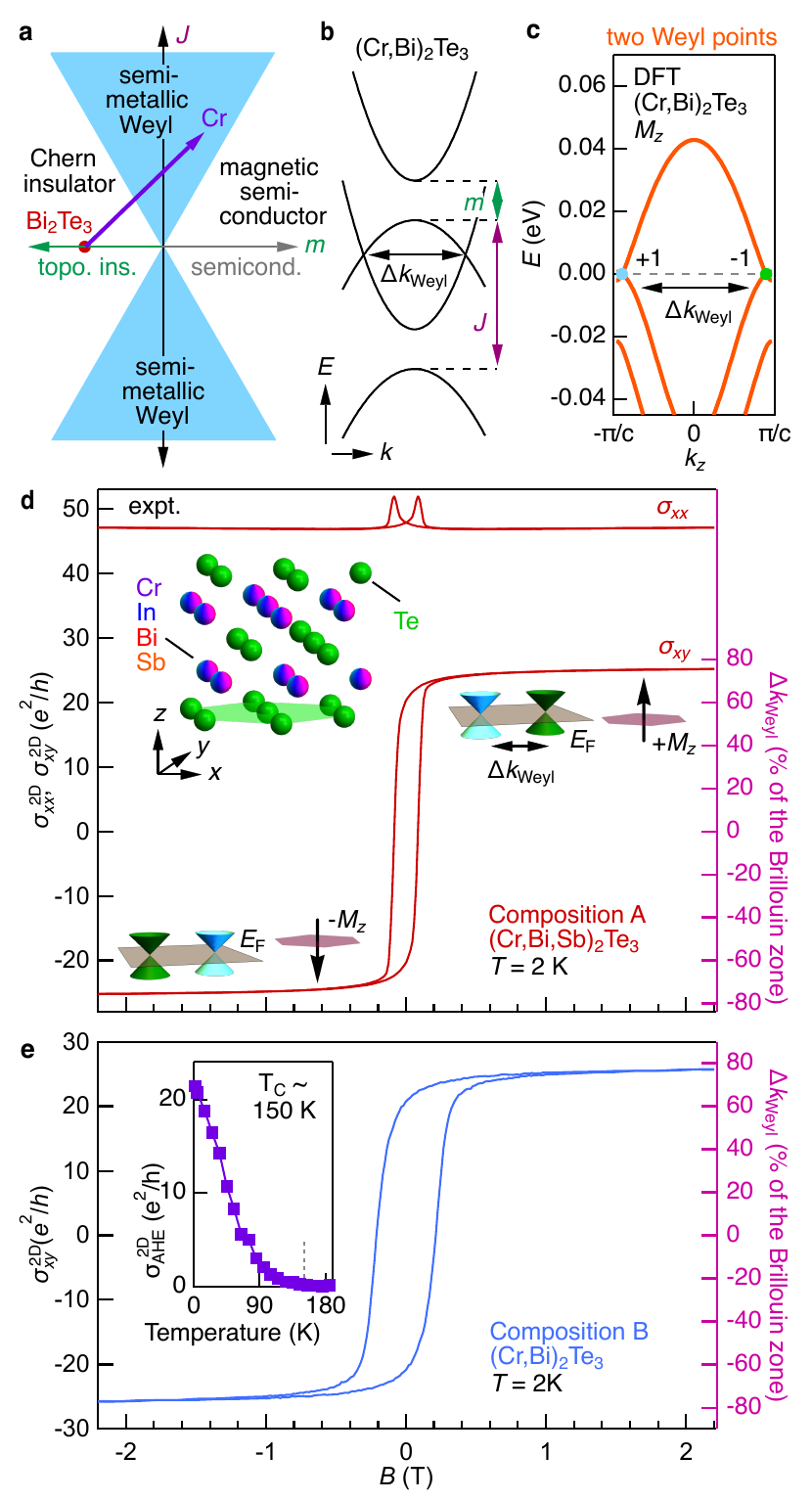}
\end{figure}

\clearpage
\begin{figure}[h]
\caption{\label{Fig1} {\bf Weyl transport semimetal in (Cr,Bi)$_2$Te$_3$.} \cpana, Phase diagram for a semiconductor with mass gap $m$ and ferromagnetic exchange splitting $J$. This diagram can be viewed as the magnetic analog to the Murakami phase diagram for a Weyl semimetal \cite{Gapless_Murakami}. \cpanb, Bi$_2$Te$_3$ is a topological insulator, with $m < 0$, $J = 0$. Cr doping introduces ferromagnetism and reduces the spin-orbit coupling (SOC), driving a band inversion and forming a semimetallic phase with two Weyl points. \cpanc, The electronic structure of (Cr,Bi)$_2$Te$_3$ under ferromagnetic order, from \ab\ calculations. The Fermi surface exhibits two Weyl points, with separation $\Delta k_{\rm Weyl} \sim 90\%$. \cpand, Longitudinal, $\sigma^{\rm 2D}_{xx}$ and Hall, $\sigma^{\rm 2D}_{xy}$ sheet conductivity of (Cr,Bi,Sb)$_2$Te$_3$, Composition A, with sample thickness $h = 100$ nm. From symmetry analysis (see main text) the Fermi surface is composed of two Weyl points, so we can directly convert the anomalous Hall effect (AHE) into a Weyl point separation, $\Delta k_{\rm Weyl} \sim 75\%$ of the bulk Brillouin zone along $k_z$. Inset: the chromium (Cr), indium (In), bismuth (Bi) and antimony (Sb) dopants randomly occupy the cation site of the Bi$_2$Te$_3$ crystal structure. \cpane, Hall sheet conductivity $\sigma^{\rm 2D}_{xy}$ for (Cr,Bi)$_2$Te$_3$, Composition B, showing $\Delta k_{\rm Weyl} \sim 77\%$. Inset: temperature dependence of the AHE, indicating a Curie temperature $T_{\rm C} \sim 150$ K.}
\end{figure}

\clearpage
\begin{figure}[h]
\centering
\includegraphics[width=13cm,trim={0in 0.2in 0in 0in},clip]{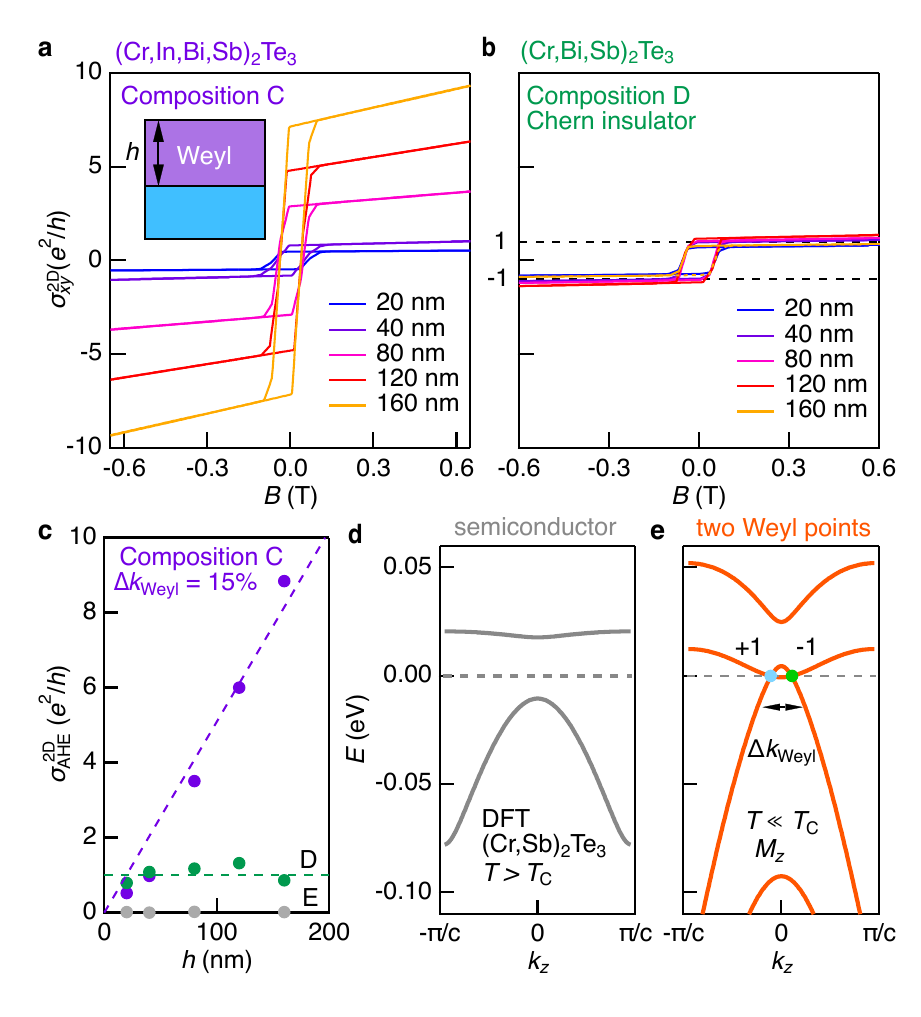}
\caption{\label{Fig2} {\bf Tunability of a semimetallic Weyl state.} Hall sheet conductivity $\sigma^{\rm 2D}_{xy}$ as a function of film thickness $h$, for \cpana, (Cr,In,Bi,Sb)$_2$Te$_3$ in the Weyl phase (Composition C) and \cpanb, (Cr,Bi,Sb)$_2$Te$_3$ in the Chern insulator phase (Composition D), at $T = 2$ K. \cpanc, Scaling of the AHE with $h$, exhibiting sharply contrasting bulk scaling (C) and surface scaling (D). Composition E: a topologically-trivial magnetic semiconductor (In,Cr,Bi,Sb)$_2$Te$_3$, exhibiting zero AHE (see \exda \ref{trivial}). \cpand, \Ab\ calculation of (Cr,Sb)$_2$Te$_3$, without magnetic order, exhibiting a semiconducting electronic structure, and \cpane, under ferromagnetism, exhibiting two Weyl points with, $\Delta k_{\rm Weyl} \sim 11\%$, roughly in agreement with $\Delta k_{\rm Weyl} \sim 15\%$ measured for `Composition C'.}
\end{figure}

\begin{figure}[h]
\centering
\includegraphics[width=14cm,trim={0in 0in 0in 0in},clip]{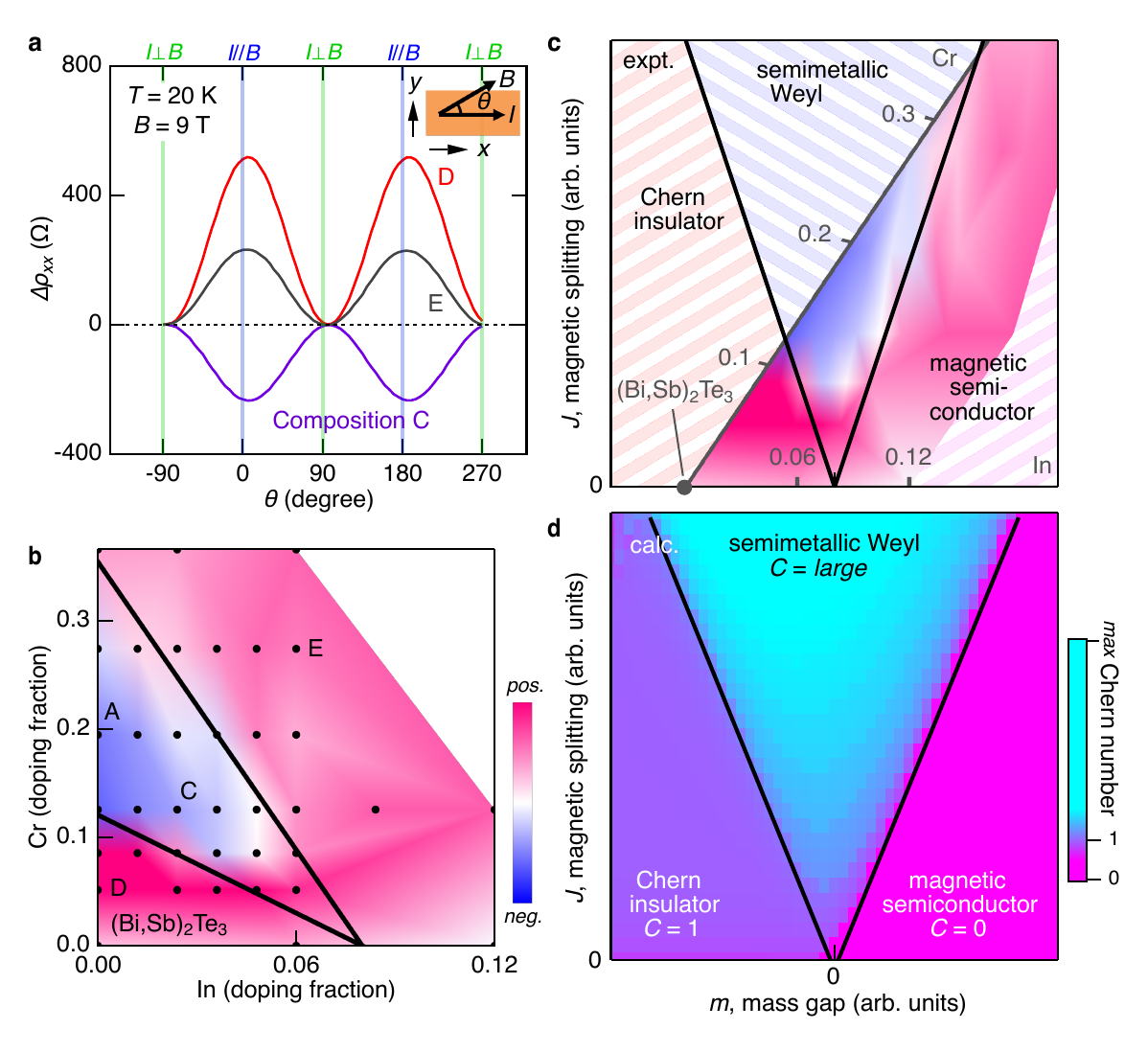}
\caption{\label{Fig3} {\bf Experimental visualization of the topological phase diagram.} \cpana, Angle-dependence of the magnetoresistance, $\Delta \rho_{xx} (\theta) = \rho_{xx} (\theta) - \rho_{xx} (\theta = 90^{\circ})$, exhibiting negative angular magnetoresistance (AMR) for the semimetallic Weyl phase (C) and positive AMR for the topological (D) and trivial (E) insulating phases. Inset: measurement geometry. \cpanb, Systematic composition dependence of the AMR across 37 films (black dots), $h = 20$ nm, with varying In and Cr content. The color map is obtained from the measured $\Delta \rho_{xx}(\theta = 0^{\circ})$ at each black dot, revealing large regions with negative (blue) and positive (red) AMR. \cpanc, The original Murakami phase diagram considers a Weyl semimetal driven by inversion-breaking \cite{Gapless_Murakami}. The magnetic analog to the Murakami diagram has ferromagnetic exchange splitting $J$, breaking time-reversal symmetry. Coordinate transformation of the color map in (\cpanb) from (In,Cr) composition axes to ($m$,$J$) axes. Note that In predominantly suppresses the spin-orbit coupling (SOC), increasing $m$ from negative to positive; Cr introduces ferromagnetism and suppresses SOC, increasing both $J$ and $m$. \cpand, Numerical calculation of the Chern number for a finite-size lattice model, exhibiting Chern insulator $C = 1$, semimetallic Weyl $C > 1$ and magnetic semiconductor $C = 0$ phases (see Methods).}
\end{figure}

\begin{figure}[h]
\centering
\includegraphics[width=17cm,trim={0in 0.2in 0in 0in},clip]{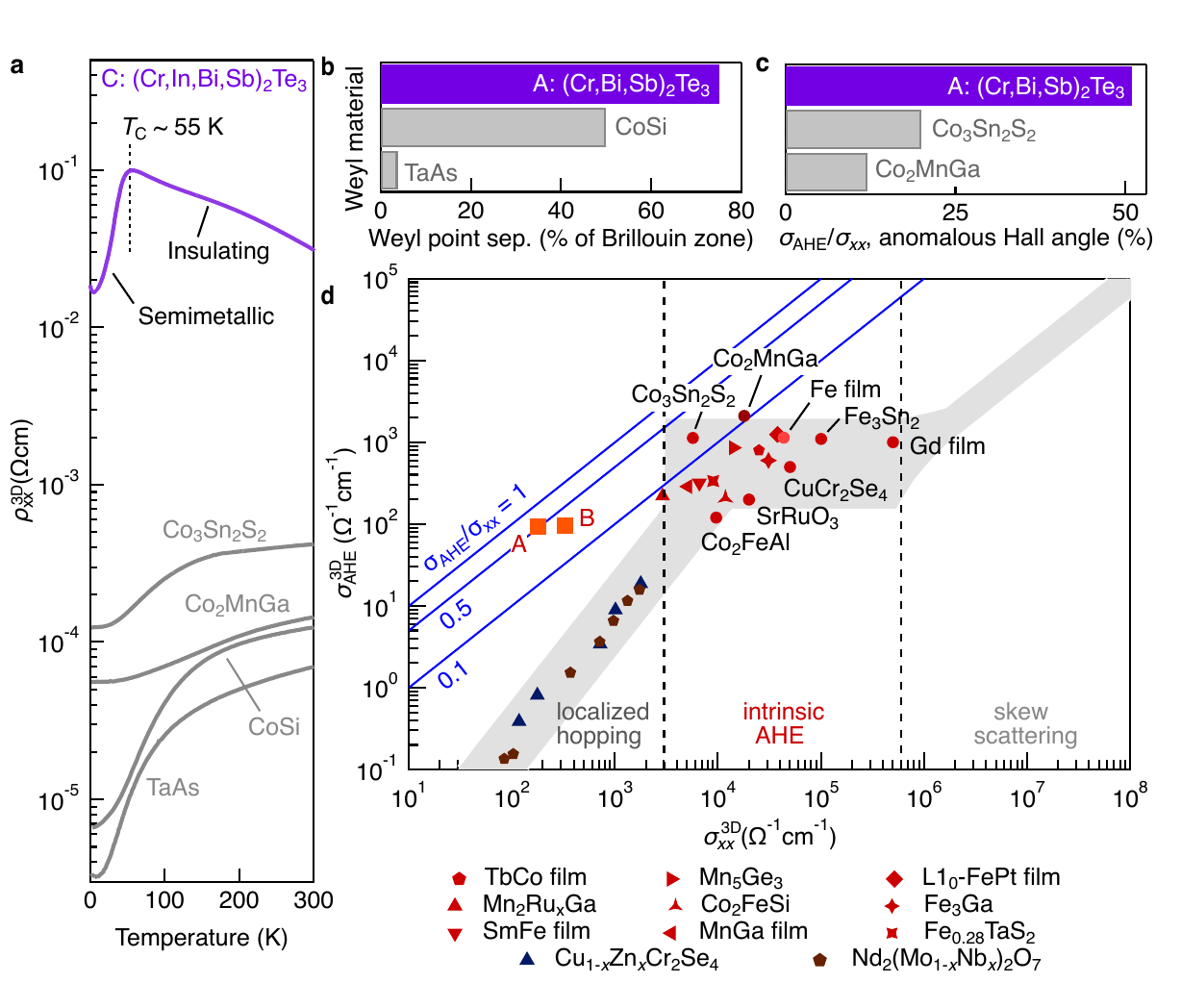}
\caption{\label{Fig4} {\bf Large Weyl point separation \& large anomalous Hall angle.} \cpana, The conventionally best-established Weyl materials exhibit metallic resistivity $\rho_{xx} (T)$ due to irrelevant, conventional electrons in the electronic structure (gray). For our semimetallic Weyl system $\rho_{xx} (T)$ is substantially enhanced, characteristic of a transport semimetal (purple, $h = 100$ nm). Our (Cr,Bi)$_2$Te$_3$ platform further exhibits \cpanb, anomalous Hall angle up to $\sigma_{\rm AHE}/\sigma_{xx} = 0.51$ and \cpanc, Weyl point separation up to $75\%$ of the Brillouin zone, substantially larger than all existing Weyl materials. \cpand, Map of $\sigma_{\rm AHE}$ vs. $\sigma_{xx}$ for many ferromagnets, illustrating the typical regimes of the localized hopping, intrinsic and skew scattering mechanisms. In the present case, (Cr,Bi)$_2$Te$_3$ shows intrinsic AHE, but lies outside of the typical intrinsic regime, due to the Weyl point Fermi surface, with enhanced anomalous Hall angle. Data points for A and B obtained from Fig. \ref{Fig1}{\bf d,e}, $T = 2$ K.}
\end{figure}

\setcounter{figure}{0}
\renewcommand{\figurename}{Extended Data Fig.}

\setcounter{table}{0}
\renewcommand{\tablename}{Extended Data Table}

\begin{figure}
\centering
\includegraphics[width=14cm,trim={0in 0in 0in 0in},clip]{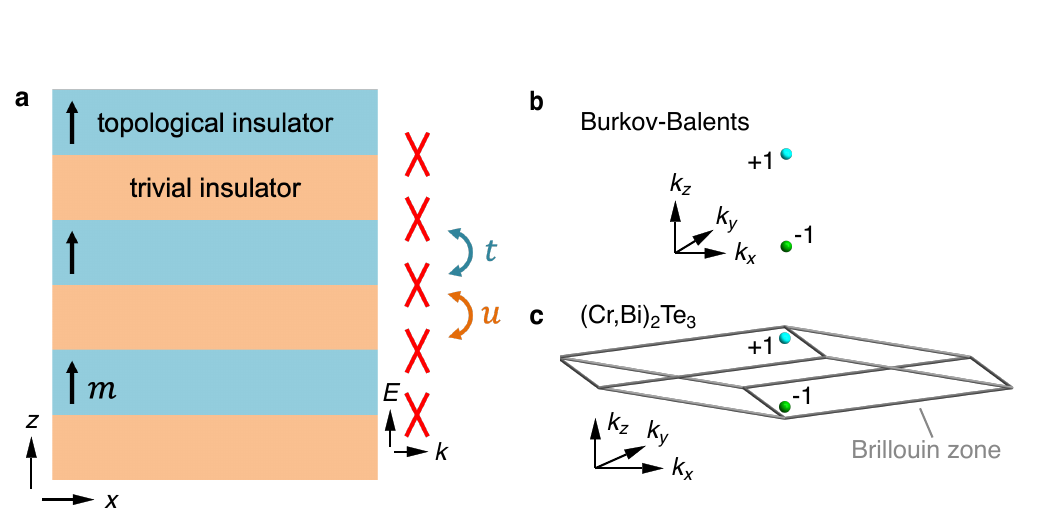}
\caption{\label{BBcartoon} {\bf Burkov-Balents proposal without a multilayer.} \cpana, The original theoretical model considers a stack of alternating layers of topological and trivial insulators hosting two-dimensional Dirac cones (red cones) at each interface. These interface states hybridize with one another with amplitude $t$ across the topological layer and $u$ across the trivial layer; there is also ferromagnetism $m$ \cite{TINI_BurkovBalents}. \cpanb, The resulting electronic structure exhibits a semimetallic Weyl phase with two Weyl points along $k_z$ (cyan, green dots). \cpanc, We circumvent the complicated multilayer structure and realize the same electronic structure as the Burkov-Balents proposal in homogeneously-doped thin films of (Cr,Bi)$_2$Te$_3$.}
\end{figure}

\clearpage

\begin{figure}
\centering
\includegraphics[width=10cm,trim={0in 0in 0in 0in},clip]{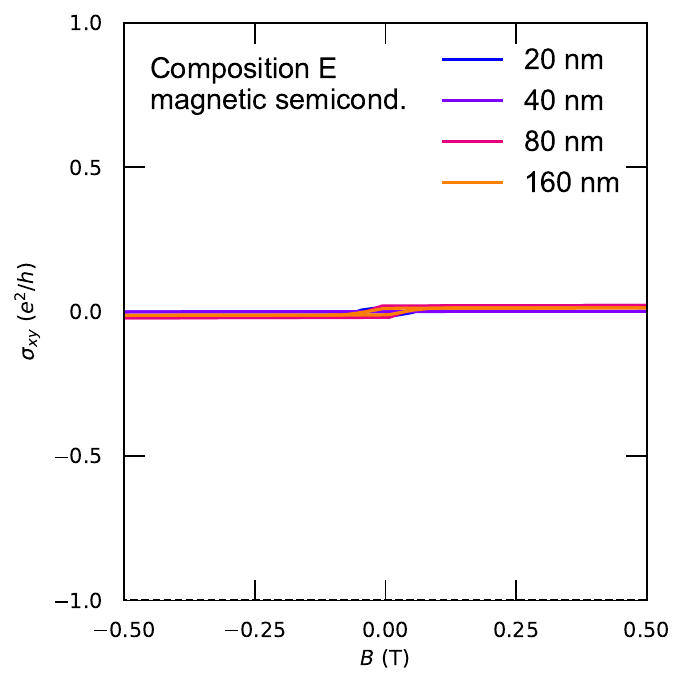}
\caption{\label{trivial} {\bf Systematics on the magnetic semiconductor.} Anomalous Hall response of `Composition E' a trivial magnetic semiconductor. The film exhibits negligible AHE $\ll 1$ $e^2/h$, for all thicknesses.}
\end{figure}

\begin{figure}
\centering
\includegraphics[width=14cm,trim={0in 0in 0in 0in},clip]{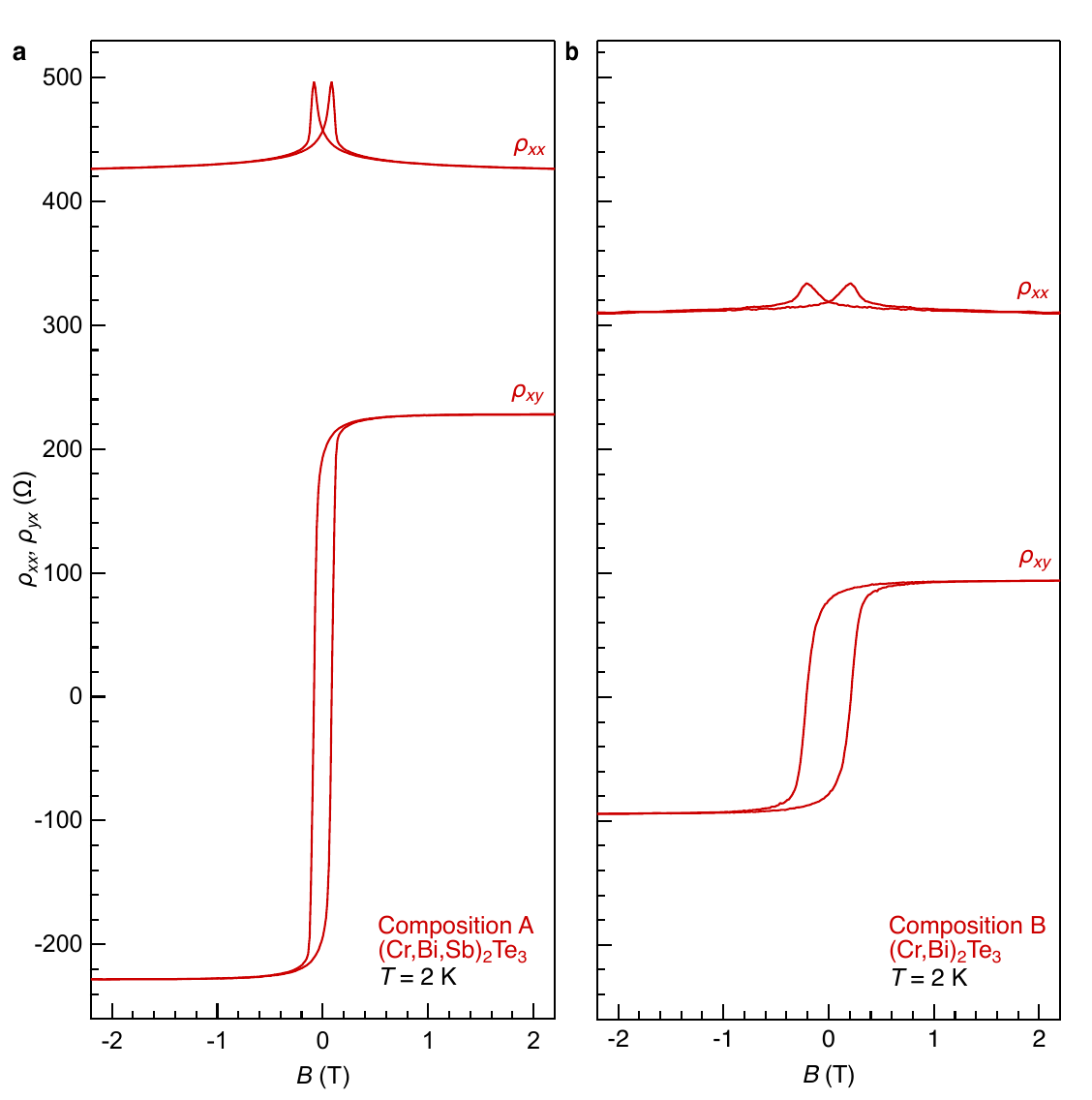}
\caption{\label{A_rho} {\bf Resistivity of Compositions A, B.} Measured $\rho_{xx}(B)$ and $\rho_{yx}(B)$ for (Cr,Bi,Sb)$_2$Te$_3$ and (Cr,Bi)$_2$Te$_3$, used to calculate the conductivity $\sigma(B)$ in main text Fig. 1d,e.}
\end{figure}

\begin{figure}
\centering
\includegraphics[width=8cm,trim={0in 0in 0in 0in},clip]{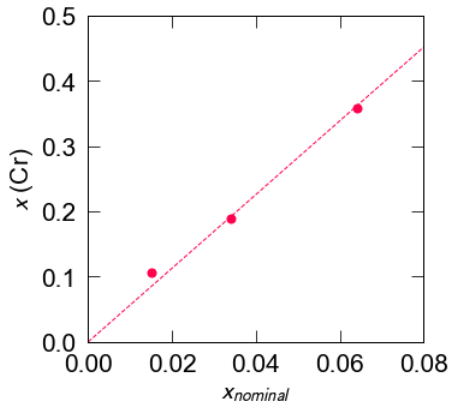}
\caption{\label{ICP} {\bf Measured Cr composition.} Chromium content $x$ for a series of (Cr,Bi,Sb)$_2$Te$_3$ films measured by inductively coupled plasma mass spectrometry (ICP-MS), plotted against the nominal chromium content $x_{\rm nominal}$ determined by beam flux pressure during film synthesis. We observe $x \sim 5.7 \times x_{\rm nominal}$.}
\end{figure}

\begin{figure}
\centering
\includegraphics[width=10cm,trim={0in 0in 0in 0in},clip]{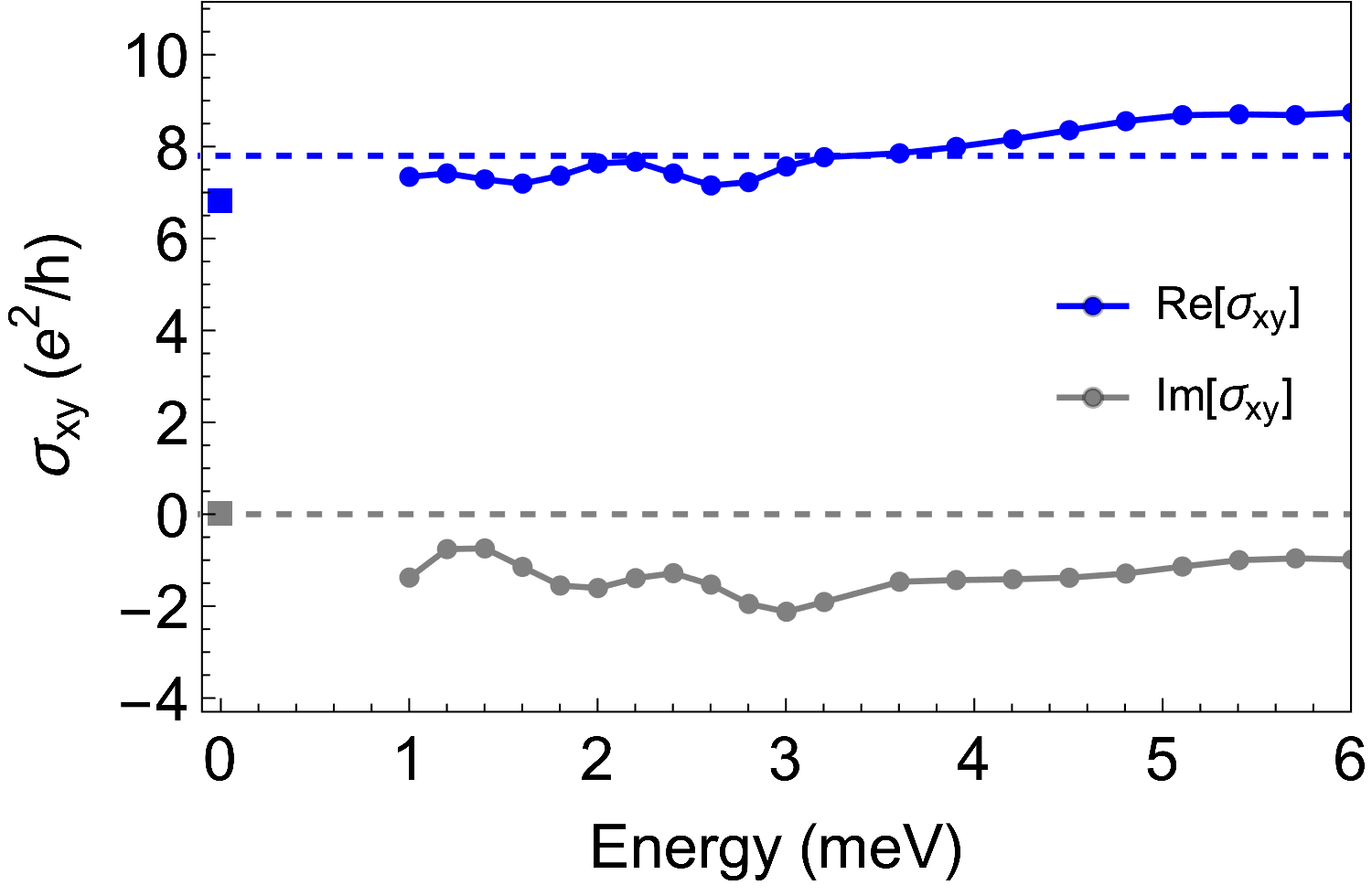}
\caption{\label{THz} {\bf Optical conductivity.} Frequency-dependent Hall conductivity in the THz range for a Weyl film, `Composition C' of thickness $h = 160$ nm. The low-frequency $\Re[\sigma_{xy}]$ is approximately constant and consistent with DC transport (square points).}
\end{figure}

\begin{figure}
\centering
\includegraphics[width=14cm,trim={0in 0.2in 0in 0in},clip]{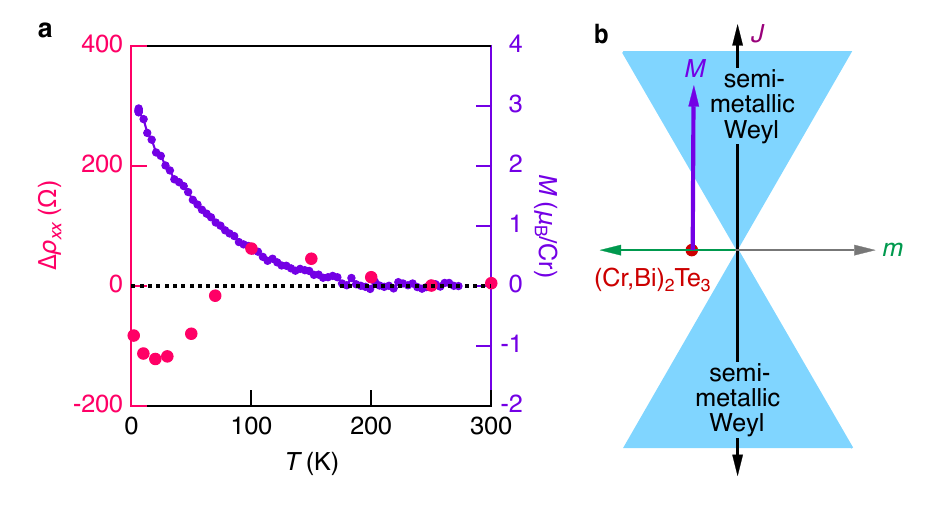}
\caption{\label{AMRtemp} {\bf Temperature dependence, angular magnetoresistance.} \cpana, Temperature dependence of the angular magnetoresistance (AMR) at $x_{\rm Cr} = 0.13$ and $x_{\rm In} = 0$ (included in Fig. \ref{Fig3}b, between `Composition C' and `Composition D'). \cpanb, The change in sign of the AMR suggests that as the magnetization $M$ develops upon cooling, the system exhibits a transition from a Chern insulator to a semimetallic Weyl phase.}
\end{figure}

\begin{figure}[h]
\centering
\includegraphics[width=15cm,trim={0in 0in 0in 0in},clip]{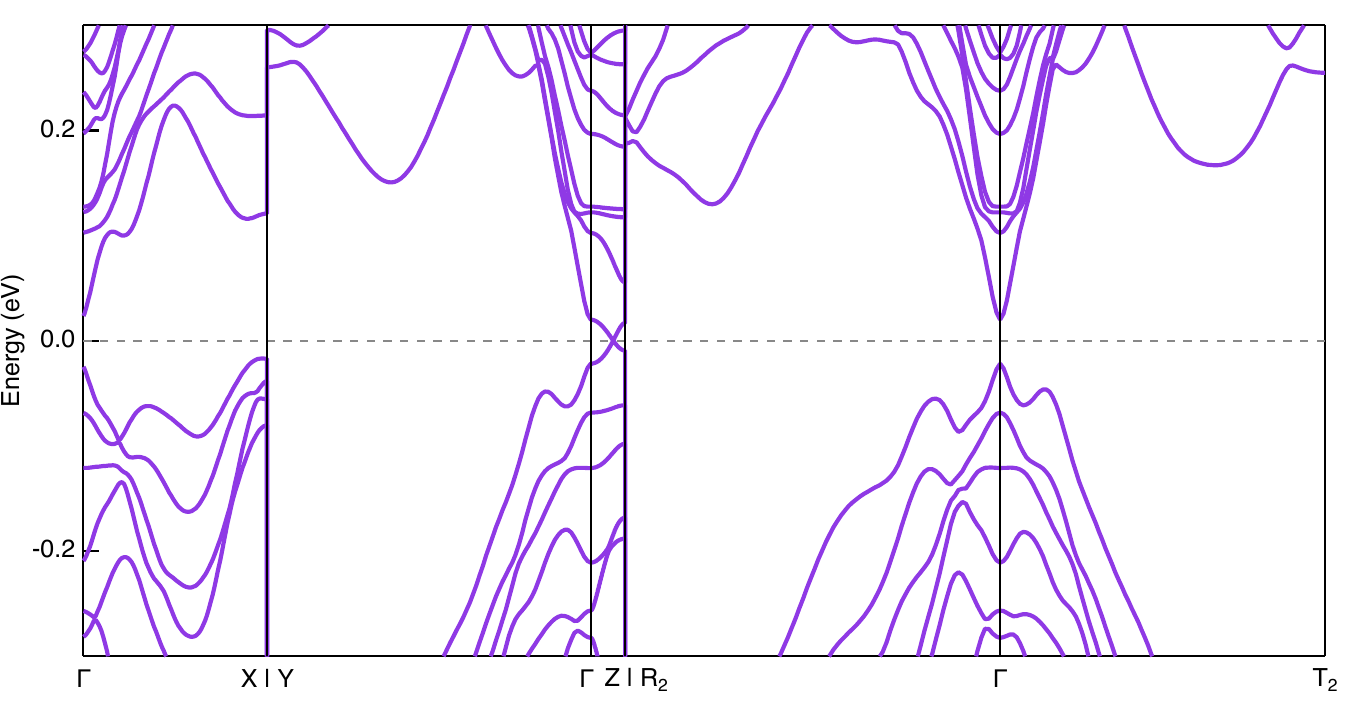}
\caption{\label{DFT_disorder} {\bf Wide-range electronic structure of Cr$_{0.25}$Sb$_{1.75}$Te$_3$ under Cr disorder.} The electronic structure was calculated for an Sb$_2$Te$_3$ supercell with 1 of 8 Sb atoms replaced by Cr dopant atoms, under multiple dopant disorder configurations. The resulting onsite potentials were averaged out. The disordered electronic structure exhibits a single pair of Weyl points along $\Gamma$-$Z$, with Weyl point separation $\Delta k_z = 48\%$, and no other bands at the Fermi level.}
\end{figure}

\clearpage

\makeatletter
\AtBeginEnvironment{tabular}{%
  \def\baselinestretch{1.1}\@currsize}%
\makeatother

% {\textbf{(Space Group)}}

% $\qquad \qquad $

\begin{table}%[htbp]\small
\begin{center}
%	\centering
%\renewcommand{\arraystretch}{1.45}
\begin{tabular}{{ m{2.5cm}<{\centering} m{2.5cm}<{\centering} m{2cm}<{\centering}  m{2cm}<{\centering} m{2cm}<{\centering} m{2cm}<{\centering} }}
		\toprule
		{\textbf{Crystal}} & {\textbf{Lattice}} & \multicolumn{4}{c}{\textbf{Atomic coordinates (fractional)}} \\
		& (${\rm \AA}$, deg.) & \textbf{Atom} & \textbf{X} & \textbf{Y} & \textbf{Z} \\
		\hline
		{Cr$_2$Sb$_4$Te$_9$} & {$a = 4.3039$} & Cr1 & 0.00000 & 0.00000 & 0.39580 \\
		$P\bar{3}m1$ & {$b = 4.3039$} & Cr2 & 0.00000 & 0.00000 & 0.60420 \\
		& {$c = 31.7773$} & Sb1 & 0.66667& 0.33333& 0.72913 \\
		& $\alpha = \beta = 90^\circ$ & Sb2 & 0.66667& 0.33333& 0.72913 \\
		& $\gamma = 120^\circ$ & Sb3 & 0.33333& 0.66667& 0.27087 \\
		&       & Sb4 & 0.33333& 0.66667& 0.06246 \\
		&       & Te1 & 0.00000& 0.00000& 0.21745 \\
		&       & Te2 & 0.00000& 0.00000& 0.78255 \\
		&       & Te3 & 0.66667& 0.33333& 0.55079 \\
		&       & Te4 & 0.33333& 0.66667& 0.44921 \\
		&       & Te5 & 0.66667& 0.33333& 0.33333 \\
		&       & Te6 & 0.33333& 0.66667& 0.66667 \\
		&       & Te7 & 0.33333& 0.66667& 0.88412 \\
		&       & Te8 & 0.66667& 0.33333& 0.11588 \\
		&       & Te9 & 0.00000& 0.00000& 0.00000 \\
		\hline
		{Cr$_2$Bi$_4$Te$_9$} & {$a = 4.43676$} & Cr1 & 0.00000 & 0.00000 & 0.60124 \\
		$P\bar{3}m1$ & {$b = 4.43676$} & Cr2 & 0.00000 & 0.00000 & 0.39876 \\
		& {$c = 31.31840$} & Bi1 & 0.33333 & 0.66667 & 0.26791 \\
		& $\alpha = \beta = 90^\circ$ & Bi2 & 0.33333 & 0.66667 & 0.06542 \\
		& $\gamma = 120^\circ$ & Bi3 & 0.66667 & 0.33333 & 0.93458 \\
		&       & Bi4 & 0.66667 & 0.33333 & 0.73209 \\
		&       & Te1 & 0.00000 & 0.00000 & 0.00000 \\
		&       & Te2 & 0.00000 & 0.00000 & 0.21308 \\
		&       & Te3 & 0.66667 & 0.33333 & 0.12026 \\
		&       & Te4 & 0.66667 & 0.33333 & 0.33333 \\
		&       & Te5 & 0.66667 & 0.33333 & 0.54641 \\
		&       & Te6 & 0.33333 & 0.66667 & 0.45359 \\
		&       & Te7 & 0.33333 & 0.66667 & 0.66667 \\
		&       & Te8 & 0.33333 & 0.66667 & 0.87974 \\
		&       & Te9 & 0.00000 & 0.00000 & 0.78692 \\
		\hline
		%\hline
\end{tabular}
\caption{Structure of (Cr,Bi)$_2$Te$_3$ and (Cr,Sb)$_2$Te$_3$, used for \ab\ calculations.}
\label{Ext_Struct}
\end{center}
\end{table}

\end{document}